\documentclass[epj,nopacs]{svjour}
\usepackage{graphicx}

\begin{document}

\title{Comment on \textquotedblleft Gain-assisted superluminal light propagation
through a Bose-Einstein condensate cavity system\textquotedblright{}}

\titlerunning{Comment on \textquotedblleft Gain-assisted superluminal light propagation...\textquotedblright{}}
\author{Bruno Macke \and Bernard S\'{e}gard \thanks{e-mail :\texttt{bernard.segard@univ-lille1.fr}}}

\authorrunning{B. Macke and B. S\'{e}gard}                  
\institute{Laboratoire de Physique des Lasers, Atomes et Mol\'{e}cules ,\\ CNRS et
Universit\'{e} de Lille, \\ 59655 Villeneuve d'Ascq, France}

\date{Published online 22 September 2016}
\abstract{
In a recent theoretical article {[}S.H. Kazemi, S. Ghanbari, M. Mahmoudi, Eur. Phys. J. D \textbf{70}, 1 (2016){]},
Kazemi et al. claim to have demonstrated superluminal light
transmission in an optomechanical system where a Bose-Einstein condensate
serves as the mechanical oscillator. In fact the superluminal propagation
is only inferred from the existence of a minimum of transmission of
the system at the probe frequency. This condition is not sufficient
and we show that, in all the cases where superluminal propagation
is claimed by Kazemi et al., the propagation is in reality
subluminal. Moreover, we point out that the system under consideration
is not minimum-phase-shift. The Kramers-Kronig relations then only
fix a lower limit to the group delay and we show that these two quantities
have sometimes opposite signs.
}
\maketitle
When the transmission of light in a given medium displays a well-marked
narrow dip at some frequency, the group velocity at this frequency
may be larger than the velocity of light in vacuum or even negative.
An ideally smooth light-pulse can then exits the medium without significant
distortion before than if it had propagated in vacuum \cite{Ref1}.
Such superluminal or fast propagation is not at odds with relativistic
causality since a given point of the output-pulse profile is not a
direct reflection of the homologous point of the incident-pulse profile
but results from the action of the medium on all the earlier part
of the incident pulse. The major challenge in such experiments is
to obtain advancements comparable to the pulse duration with moderate
distortion. Convincing experiments have been performed in the 1980s
\cite{Ref2,Ref3} in media with a narrow absorption line. Unfortunately,
superluminal propagation is then accompanied by strong absorption.
This inconvenience is overcome by using a medium with a doublet of
gain lines \cite{Ref4,Ref5} and a minimum of transmission between
them. Significant advancements have been evidenced in an atomic vapor
with this arrangement \cite{Ref6}. A comprehensive review on fast
light in atomic media can be found in \cite{Ref7}. Experiments involving
four wave mixing are reported in \cite{Ref8}.

 Superluminal or subluminal propagation can only be demonstrated by a determination of the group
delay and one should not hastily conclude from what it precedes that
every gain system with a dip in its transmission curve will be superluminal.
This extrapolation is unfortunately made in a recent theoretical article
\cite{Ref9} whose authors claim to have evidenced superluminal propagation
by giving this sole argument. The system under consideration is an
optomechanical device consisting in a high-quality optical cavity
containing a Bose-Einstein condensate (BEC) of Rubidium atoms which
serves as the mechanical oscillator \cite{Ref10}. It is submitted
to a strong pump field (continuous wave) and to a weak probe field.
In a frame rotating at the pump angular-frequency, the transfer function
for the probe field reads as \cite{Ref9}\footnote{Some typo errors in \cite{Ref9} are corrected in the present comment.} 
\begin{equation}
H(\omega)=1-\frac{1+if\left(\omega\right)}{\kappa/2-2f\left(\omega\right)\Delta_{c}+i\left(\Delta_{c}-\omega\right)}(\frac{\kappa}{2})\label{eq:1}
\end{equation}
with
\begin{equation}
f\left(\omega\right)=\frac{2\omega_{m}ng^{2}}{\left[\kappa/2-i\left(\omega+\Delta_{c}\right)\right]\left[\omega_{m}^{2}-\omega^{2}-i\omega\gamma_{m}\right]}\label{eq:2}
\end{equation}
In these expressions, $\kappa$ ($\gamma_{m}$ ) is the damping rate
of the cavity (the mechanical oscillator), $\Delta_{c}$ is a frequency
related to the cavity detuning and is assumed equal to the resonance
frequency $\omega_{m}$ of the mechanical oscillator in all the simulations,
$n=E_{pu}^{2}/\left(\Delta_{c}^{2}+\kappa^{2}/4\right)$ where $E_{pu}$
is a parameter proportional to the amplitude of the pump field, $g=g_{0}^{2}\sqrt{N}/\left(2\Delta_{a}\sqrt{2}\right)$
where $g_{0}$ is the elementary atom-photon coupling constant, $N$
is the number of atoms in the BEC and $\Delta_{a}$ is the detuning
of the pump from the frequency of the relevant atomic line. We denote
in the following $T\left(\omega\right)=\left|H\left(\omega\right)\right|^{2}$
the intensity transmission of the system and $\varphi\left(\omega\right)=\arg\left[H\left(\omega\right)\right]$
the phase shift induced by the system.

\begin{figure}[h]
\begin{centering}
\includegraphics[width=85mm]{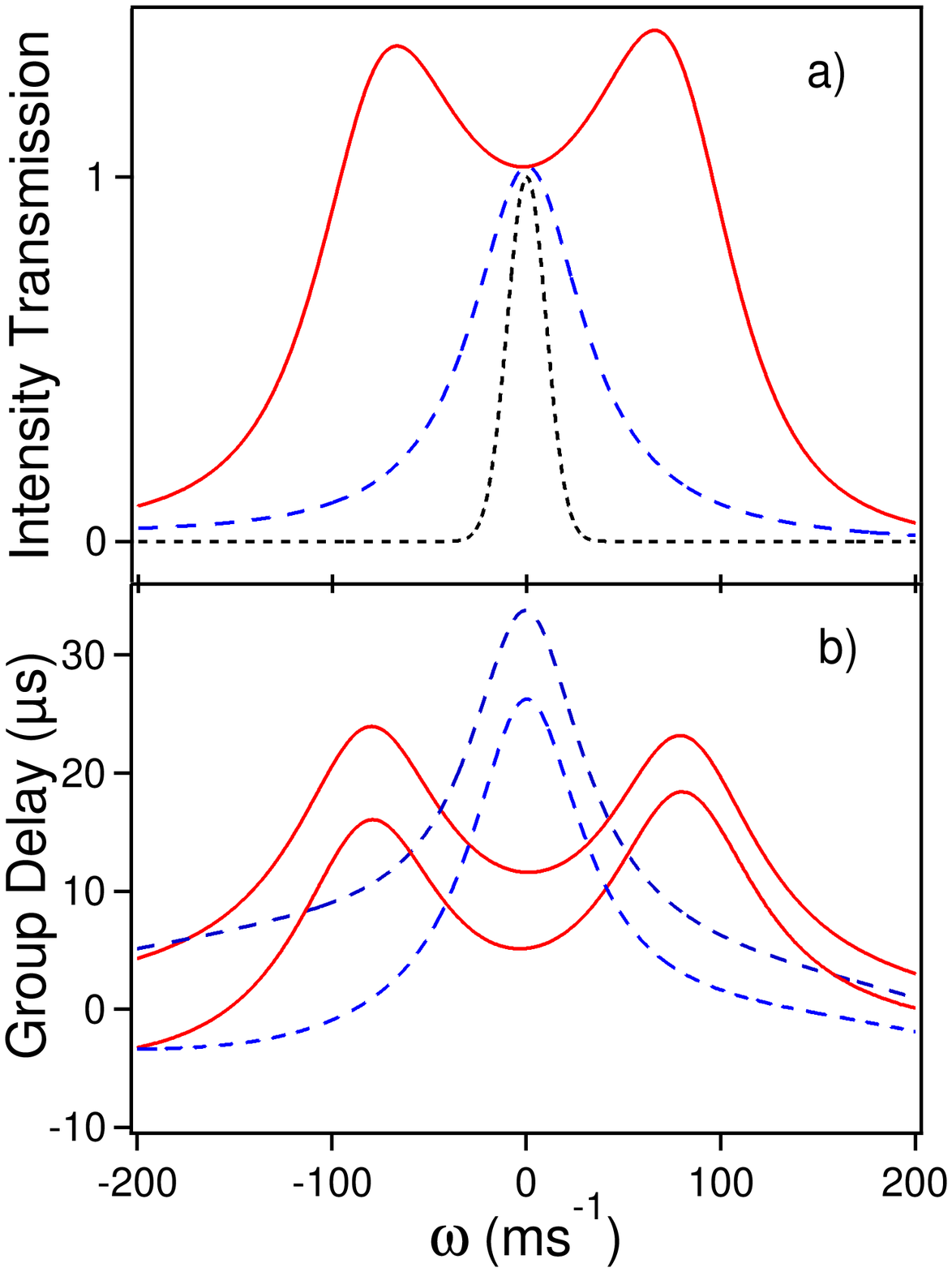} 
\par\end{centering}
\caption{a) Intensity transmission $T(\omega)$ of the optomechanical system
as a function of $\omega$. Used parameters are: $N=1.2\times10^{5}$,
$g_{0}=2\pi\times10.9\:\mathrm{MHz}$, $\Delta_{a}=2\pi\times32\:\mathrm{GHz}$,
$\kappa=2\pi\times1.3\:\mathrm{MHz}$, $\Delta_{c}=\omega_{m}=2\pi\times15.2\:\mathrm{kHz}$,
$E_{pu}=2\pi\times100\:\mathrm{kHz}$, $\gamma_{m}=2\pi\times15\:\mathrm{kHz}$
(solid line) and $\gamma_{m}=2\pi\times45\:\mathrm{kHz}$ (dashed
line). The dotted line is the power spectrum of the incident pulse
considered in Figure.\ref{fig:Temporal-response}. b) Corresponding group
delays. The upper curves give the exact group delays $\tau_{g}(\omega)=\frac{d}{d\omega}\arg\left[H\left(\omega\right)\right]$
and the lower curves give the group delays $\tau_{KK}\left(\omega\right)$
derived from the Kramers-Kronig relations. \label{fig1}}
\end{figure}

In order to evidence that the existence of a minimum of the intensity
transmission does not entail superluminal propagation we have determined
the corresponding group delay $\tau_{g}\left(\omega\right)=\frac{d}{d\omega}\varphi\left(\omega\right)$
in all the cases considered in \cite{Ref9}. As a representative example,
our Figure.\ref{fig1} shows the dependence of $T\left(\omega\right)$
and $\tau_{g}\left(\omega\right)$ as functions of $\omega$ for two
different values of the damping rate $\gamma_{m}$ of the mechanical
oscillator (BEC).
The parameters are those considered to obtain Figure.4 in \cite{Ref9}.
We see that, for $\gamma_{m}=2\pi\times15\:\mathrm{kHz}$ (solid line),
the transmission has a minimum for $\omega=0$ (probe frequency equal
to the pump frequency) but that the propagation remains subluminal
{[}$\tau_{g}(0)>0$ {]} contrary to the claim of Kazemi et al. \cite{Ref9}.
Compared to the result obtained for $\gamma_{m}=2\pi\times45\:\mathrm{kHz}$
(dashed line) where the transmission has a maximum for $\omega=0$
, it appears that the effect of a minimum of transmission is to significantly
reduce the group delay \emph{without changing its sign} (no time-advancement).
In the present case, we get $\tau_{g}(0)=33.8\:\mathrm{\mu s}$ for
$\gamma_{m}=2\pi\times45\:\mathrm{kHz}$ and $\tau_{g}(0)=11.6\:\mathrm{\mu s}$
for $\gamma_{m}=2\pi\times15\:\mathrm{kHz}$ . Figure 2 shows the
intensity profiles of the pulses transmitted by the system when it
is subjected to an incident Gaussian pulse with a carrier frequency
equal to the pump frequency. 
\begin{figure}[h]
\begin{centering}
\includegraphics[width=85mm]{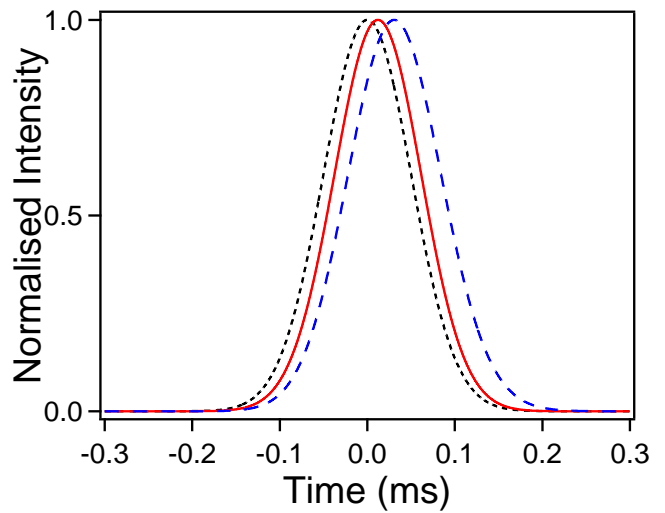} 
\par\end{centering}
\caption{Temporal response of the system to an incident Gaussian pulse of carrier
frequency equal to the pump frequency and of duration $\tau_{p}=100\:\mathrm{\mu s}$.
(half-width at $1/e$ of the envelope). The figure shows the normalized
intensity-profiles of the incident pulse (dotted line), of the pulse
transmitted for $\gamma_{m}=2\pi\times15\:\mathrm{kHz}$ (solid line) and for
$\gamma_{m}=2\pi\times45\:\mathrm{kHz}$ (dashed line). The other parameters are those of Figure.\ref{fig1}.\label{fig:Temporal-response}}
\end{figure}
The intensity profile of the incident pulse is given for reference
(dotted line). Its duration $\tau_{p}$ (half-width at $1/e$ of its
envelope) has been taken equal to $100\:\mathrm{\mu s}$ in order
that $T(\omega)$ and $\tau_{g}(\omega)$ do not vary too considerably
over the width of its power spectrum (shown in dotted line in Figure.\ref{fig1}a).
The pulse distortion remains then moderate and the delay $t_{m}$
of the pulse maximum is close to the group delay \cite{Ref5}. We
get $t_{m}=30.7\:\mathrm{\mu s}$ for $\gamma_{m}=2\pi\times45\:\mathrm{kHz}$
and $t_{m}=12.1\:\mathrm{\mu s}$ for $\gamma_{m}=2\pi\times15\:\mathrm{kHz}$
. Similar results are obtained in all the cases where Kazemi et al.
\cite{Ref9} predict superluminal propagation. Even in the conditions
of their Figure.5 where the gain dynamics is particularly large, we find
that \emph{the group delay remains positive}, namely $\tau_{g}(0)=5.6\:\mathrm{\mu s}$.

Kramers-Kronig relations are invoked in \cite{Ref9} to associate
superluminal propagation with a minimum of the medium transmission.
As extensively shown in \cite{Ref11,Ref12}, these relations only
give the exact group delay when the system is minimum-phase-shift
\footnote{Note that, even in minimum-phase-shift systems as are the purely propagative
systems, the group delay may be positive at a frequency for which
the transmission presents a minimum. See for example Section.4 in \cite{Ref5}. }. The phase-shift and the group delay then reads as $\varphi\left(\omega\right)=\varphi_{KK}\left(\omega\right)$
and $\tau_{g}\left(\omega\right)=\tau_{KK}\left(\omega\right)=\frac{d}{d\omega}\varphi_{KK}\left(\omega\right)$
where $\varphi_{KK}\left(\omega\right)$ is the Hilbert transform
of $\ln\left[\left|H\left(\omega\right)\right|\right]$. It appears
that, in all the cases considered in \cite{Ref9}, the transfer function
$H\left(\omega\right)$ has a zero $\tilde{\omega_{0}}$ in the upper
half-plane of the complex plane and thus that the system is not minimum-phase-shift. In this case, $\tau_{KK}\left(\omega\right)$ only fixes
a lower limit to the exact group delay. The transfer function can
then be written as $H\left(\omega\right)=H_{MP}\left(\omega\right)H_{AP}\left(\omega\right)$
where $H_{MP}\left(\omega\right)$ and $H_{AP}\left(\omega\right)$
are respectively associated with a minimum-phase-shift system and
with an all-pass system, with $\left|H_{MP}\left(\omega\right)\right|=\left|H\left(\omega\right)\right|$
and $\left|H_{AP}\left(\omega\right)\right|=1$. $H_{AP}\left(\omega\right)$
is a so-called Blaschke product \cite{Ref11} which, in the present
case, is reduced to
\begin{equation}
H_{AP}\left(\omega\right)=\frac{1-\omega/\tilde{\omega}_{0}}{1-\omega/\tilde{\omega}_{0}^{\ast}}\label{eq:3}
\end{equation}
This term is responsible of additional contributions $\varphi_{AP}\left(\omega\right)$
to the phase shift $\varphi_{KK}\left(\omega\right)$ and $\tau_{AP}\left(\omega\right)$
to the group delay $\tau_{KK}\left(\omega\right)$ . These contributions
read as
\begin{equation}
\varphi_{AP}\left(\omega\right)=\arg\left[H_{AP}\left(\omega\right)\right]=-2\tan^{-1}\left[\frac{\mathrm{Im}\left(\omega/\tilde{\omega}_{0}\right)}{1-\mathrm{Re}\left(\omega/\tilde{\omega}_{0}\right)}\right]\label{eq:4}
\end{equation}
\begin{equation}
\tau_{AP}\left(\omega\right)=\frac{d}{d\omega}\varphi_{AP}\left(\omega\right)=-\frac{2\mathrm{Im}\left(1/\tilde{\omega}_{0}\right)}{1+\left|\omega/\tilde{\omega}_{0}\right|^{2}-\mathrm{2Re}\left(\omega/\tilde{\omega}_{0}\right)}\label{eq:5}
\end{equation}
$\mathrm{Im}\left(\tilde{\omega}_{0}\right)$ being positive, $\mathrm{Im}\left(1/\tilde{\omega}_{0}\right)$
is negative and $\tau_{AP}\left(\omega\right)$ is always positive
as expected. For $\omega=0$, $\tau_{AP}\left(\omega\right)$ takes
the simple form $\tau_{AP}\left(0\right)=-\mathrm{Im}\left(1/\tilde{\omega}_{0}\right)$
. In the conditions of Figure.\ref{fig1}, we get $\tilde{\omega}_{0}=-0.127+0.244i\:\mathrm{\mu s^{-1}}$
($-0.115+0.200i\:\mathrm{\mu s^{-1}}$ ) for $\gamma_{m}=2\pi\times15\:\mathrm{kHz}$
($\gamma_{m}=2\pi\times45\:\mathrm{kHz}$). In both cases the difference
between the exact group delay and that given by the Kramers-Kronig
relations is everywhere positive and perfectly reproduced by Eq.(\ref{eq:5}).
For $\omega=0$, we get in particular $\tau_{AP}=6.45\:\mathrm{\mu s}$
($7.51\:\mathrm{\mu s}$), $\tau_{KK}=5.16\:\mathrm{\mu s}$ ($26.3\:\mathrm{\mu s}$)
and $\tau_{g}=11.6\:\mathrm{\mu s}$ ($33.8\:\mathrm{\mu s}$), with
$\tau_{g}=\tau_{KK}+\tau_{AP}$ as expected.

\begin{figure}[h]
\begin{centering}
\includegraphics[width=85mm]{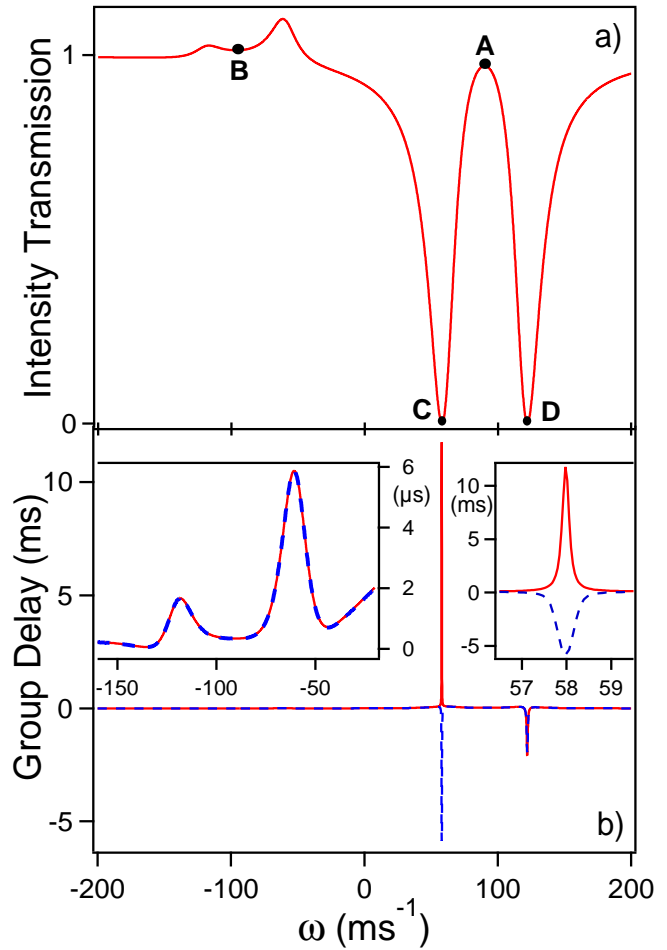} 
\par\end{centering}
\caption{a) Intensity transmission $T(\omega)$ of the optomechanical system
as a function of $\omega$. Used parameters are now : $\Delta_{a}=2\pi\times2.77\:GHz$
, $\gamma_{m}=2\pi\times400\:Hz$ and $\kappa=2\pi\times7.2\:kHz$.
The parameters not defined in \cite{Ref9} have been chosen in order
to reproduce the Figure.6 of this article. b) Corresponding group delays.
The solid line gives the exact group delay $\tau_{g}(\omega)=\frac{d}{d\omega}\arg\left[H\left(\omega\right)\right]$
and the dashed line gives the group delays $\tau_{KK}\left(\omega\right)$
derived from the Kramers-Kronig relations. The insets are enlargements in the spectral regions around the frequencies  $\omega_{B}$  and  $\omega_{C}$ .\label{figTransandDelayFig6}}
\end{figure}

Arrived to this point, it is worth remarking that significant differences
between the exact group delay $\tau_{g}$ and the group delay $\tau_{KK}$
derived from the Kramers-Kronig relations are not specific to the
system considered in \cite{Ref9}. They are often obtained in optical
systems involving mirrors and/or polarizers. See, e.g., \cite{Ref13,Ref14,Ref15}.
These two quantities may even have opposite signs. This phenomenon
marginally occurs in the conditions of our Figure.\ref{fig1}. On the
left of Figure.\ref{fig1}b, we actually see that $\tau_{g}>0$ (time-delay)
whereas the Kramers-Kronig relations predict a time advancement ($\tau_{KK}<0$).
Much more spectacular effects are shown on Figure.\ref{figTransandDelayFig6}.
Figure \ref{figTransandDelayFig6}a shows the transmission $T\left(\omega\right)$
as a function of $\omega$ obtained for parameters enabling us to
reproduce the transmission curve given Figure.6 in \cite{Ref9}. For these parameters, the transfer function of the system is again not minimum-phase-shift.
Figure.\ref{figTransandDelayFig6}b shows the corresponding functions $\tau_{g}\left(\omega\right)$ (solid
line) and $\tau_{KK}\left(\omega\right)$ (dashed line). Denoting
$\omega_{A,B,C,D}$ the frequencies corresponding to the points $A,B,C,D$
in Figure.\ref{figTransandDelayFig6}a, we note that $\tau_{g}\left(\omega\right)$
and $\tau_{KK}\left(\omega\right)$ are everywhere equal or very close,
except for $\omega\approx\omega_{C}$. The inset on the left confirms
that for $\omega\approx\omega_{B}$ the propagation is not superluminal
contrary to the claim of Kazemi et al. \cite{Ref9}. As expected,
the propagation is subluminal for $\omega\approx\omega_{A}$ but the
corresponding group delay is only $22.4\:\mathrm{\mu s}$ (not visible
at the figure scale). The most interesting features are observed for
$\omega\approx\omega_{C}$ and for $\omega\approx\omega_{D}$, evidencing
that very similar transmission profiles with a well-marked minimum
can lead to quite different group delays\footnote{Analogous phenomena have been observed in optical system consisting
in a photonic crystal \cite{Ref14} or a birefringent fibre \cite{Ref15}
placed between two polarizers. These papers report in particular \emph{experimental
evidence} that subluminal propagation can be associated with a well
marked-dip in the transmission curve of a non-minimum-phase-shift
system.}. For $\omega=\omega_{C}$, the inset on the right shows that the
group delay has a very large positive value whereas an irrelevant
application of the Kramers-Kronig relations predict a group advancement
($\tau_{KK}<0$ ). For $\omega=\omega_{D}$ on the contrary, $\tau_{g}=\tau_{KK}<0$ and a
significant group advancement is obtained as currently expected at
a transmission minimum. It should be noted that large group delay
or advancement are both paid at the price of a dramatically weak transmission.
\begin{figure}[h]
\begin{centering}
\includegraphics[width=85mm]{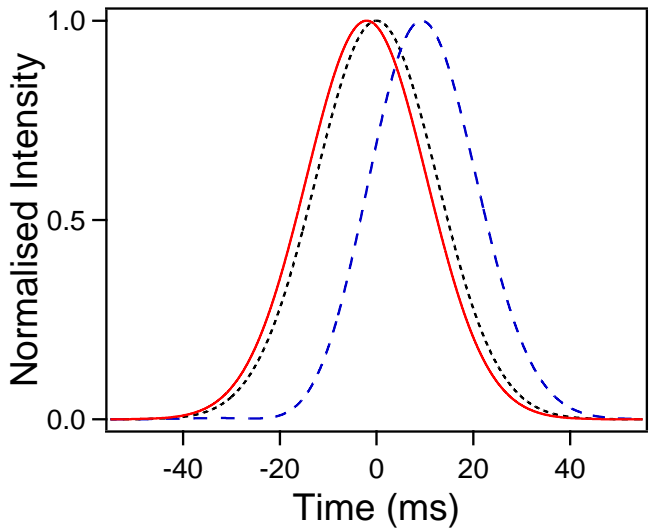} 
\par\end{centering}

\caption{Normalized intensity-profile of the pulses transmitted by the system
subjected to an incident Gaussian pulse of duration $\tau_{p}=25\:\mathrm{ms}$
with a carrier frequency detuned from the pump frequency by $\omega_{c}$
(dashed line, subluminal case) and $\omega_{D}$ (solid line, superluminal
case). The system parameters are those of Figure.\ref{figTransandDelayFig6}.
The profile of the incident pulse is given for reference (dotted line).
In the subluminal case the maximum of the output pulse is delayed
by $9.3\:\mathrm{ms}$ whereas $\tau_{g}=11.8\:\mathrm{ms}$.
In the superluminal case, this maximum is advanced by $2.06\:\mathrm{ms}$
whereas $\tau_{g}=-2.08\:\mathrm{ms}$. The corresponding absolute
intensities are respectively $1.6\times 10^{4}$ and $660$ times weaker than
that of the incident pulse.\label{TemporalFig6}}
\end{figure}
Figure.\ref{TemporalFig6} shows the intensity profiles of the
transmitted pulses when the system is subjected to incident Gaussian
pulses with a carrier frequency detuned from the pump frequency by
$\omega_{C}$ (dashed line) or by $\omega_{D}$ (solid line). The
intensity profile of the incident pulse is given for reference (dotted
line). Its duration $\tau_{p}$ has been taken as large as $25\:\mathrm{ms}$
to avoid significant pulse-distortion.

To summarize, the inconsistent claims of superluminal propagation
made in \cite{Ref9} originate in two misconceptions, firstly that
the existence of a minimum of transmission \emph{always} implies superluminal
propagation and, secondly, that the group delays can \emph{always} be derived
from the Kramers-Kronig relations whereas the latter only give a lower
limit to this delay when the system under consideration is not minimum-phase-shift
(as are numerous optical systems). Surprisingly enough, these misconceptions
are widespread in the optics community. The present comment is expected
to bring some clarification on this subject.

\section*{Author contribution statement }
The two authors contributed equally to the article.

This work has been partially supported by Ministry of Higher Education
and Research, Nord-Pas de Calais Regional Council and European Regional
Development Fund (ERDF) through the Contrat de Projets \'{E}tat-R\'{e}gion
(CPER) 2015\textendash 2020, as well as by the Agence Nationale de
la Recherche through the LABEX CEMPI project (ANR-11-LABX-0007).

\end{document}